\documentclass[aps,pre,preprint,groupedaddress]{revtex4}

\usepackage{graphicx}
\usepackage{latexsym}
\usepackage{amsmath,amssymb}

\begin{document}

%\preprint{}

\title{
Multi-criticality of the three-dimensional Ising
model with plaquette interactions:
An extension of Novotny's transfer-matrix formalism
}

\author{Yoshihiro Nishiyama}
%\thanks{}
\affiliation{Department of Physics, Faculty of Science,
Okayama University, Okayama 700-8530, Japan.}

\date{\today}

\begin{abstract}
Three-dimensional Ising model with the plaquette-type
(next-nearest-neighbor and four-spin) interactions
is investigated numerically.
This extended Ising model, the so-called gonihedric model,
was introduced by Savvidy and Wegner as a discretized version
of the interacting (closed) surfaces without surface tension.
The gonihedric model is notorious for its slow relaxation to
the thermal equilibrium (glassy behavior), which deteriorate
the efficiency of the Monte Carlo sampling.
We employ the transfer-matrix (TM) method, implementing
Novotny's idea, which enables us to treat arbitrary
number of spins $N$ for one TM slice even in three dimensions.
This arbitrariness admits systematic finite-size-scaling analyses.
Accepting the extended parameter space by Cirillo and co-worker,
we analyzed the (multi) criticality of the gonihedric model for $N \le 13$.
Thereby, we found that, as first noted by Cirillo and co-worker 
analytically (cluster-variation method),
the data are well described by the multi-critical (crossover) scaling theory.
That is, the previously reported nonstandard criticality for the
gonihedric model is reconciled with a crossover exponent and
the ordinary three-dimensional-Ising universality class.
We estimate the crossover exponent and the correlation-length 
critical exponent at the multi-critical point 
as $\phi=0.6(2)$ and $\dot{\nu}=0.45(15)$, respectively.
\end{abstract}

%\keywords{}

\maketitle

\section{\label{section1}Introduction}

Study on surfaces spans a wide variety of subjects 
ranging
from biochemistry to high-energy physics 
\cite{Nelson89,Nelson96}, leading to a very active
area of research.
In particular, the problem of
interacting surface gas \cite{Horn83,Maritan84,Karowski85,Karowski86} is of fundamental significance.
The Savvidy-Wegner (gonihedric) model 
\cite{Savvidy94,Savvidy94b,Savvidy94c,Savvidy96,Cappi93} 
describes the interacting closed
surfaces without surface tension.
The surfaces are discretized in such a way that they are
embedded in the three-dimensional cubic lattice,
and the surface faces consist of plaquettes.
The gonihedric model was introduced as a lattice-regularized version
of the string field theory \cite{Ambartzumian92}. 
However, recent developments dwell on the case
of three dimensions, aiming a potential applicability
to microemulsions.

The gonihedric model admits a familiar representation
in terms of the Ising-spin variables $\{ S_i \}$ through the duality
transformation; namely, the plaquette surfaces are
regarded as the magnetic-domain interfaces.
To be specific,
the Hamiltonian is given by the following form;
\begin{equation}
\label{gonihedric}
  H= J_1 \sum_{\langle i,j \rangle} S_i S_j 
   +J_2 \sum_{\langle \langle i, j \rangle\rangle} S_i S_j 
   +J_3 \sum_{[i,j,k,l]} S_i S_j S_k S_l
 ,
\end{equation}
with finely tuned coupling constants,
$J_1=-2 \kappa$, 
$J_2=\kappa / 2$ and 
$J_3=-(1-\kappa) / 2$.
The Ising spins $S_i = \pm 1$ are placed at the 
cubic-lattice points in three dimensions, and
the summations 
$\sum_{\langle i,j\rangle}$,
$\sum_{\langle\langle i,j \rangle\rangle}$, and
$\sum_{[i,j,k,l]}$
run over all possible nearest-neighbor pairs, 
next-nearest-neighbor (plaquette diagonal) spins, and
round-a-plaquette spins, respectively.
The interfacial energy $E$ of the gonihedric model is given by the
formula $E = n_2 + 4 \kappa n_4$, where
$n_2$ is the number of links where two plaquettes meet at a right angle
(folded-link length)
and $n_4$ is the number of links where four plaquettes meet at right angles
(self-intersection-link length).
Namely,
the surfaces are subjected to a bending elasticity 
with a fixed strength, and the 
the self-avoidance is controlled by the parameter $\kappa$.
We notice that the interfacial energy lacks the surface-tension
term.

\begin{figure}
\includegraphics{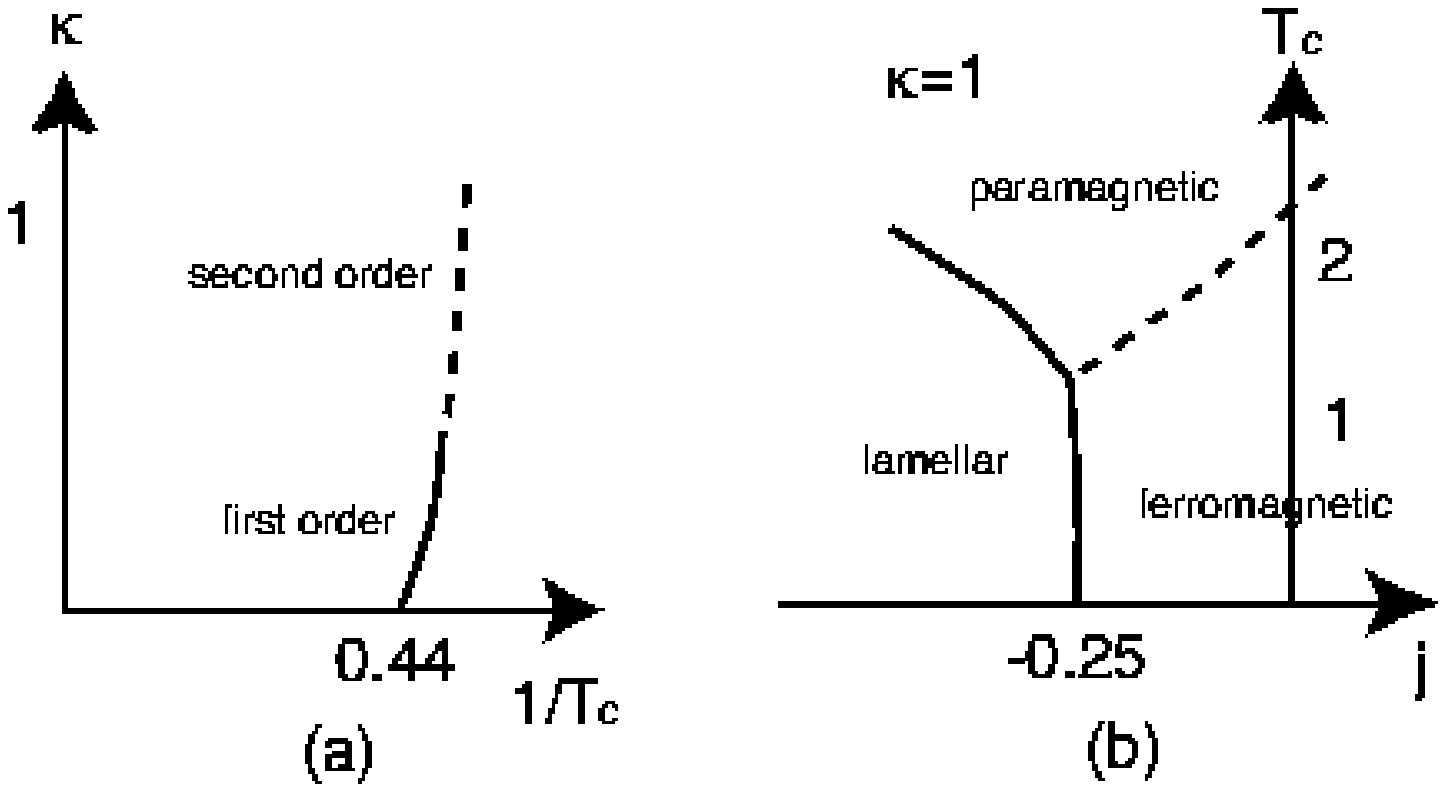}%
\caption{\label{figure1}
(a) A schematic phase diagram for the gonihedric model
(\ref{gonihedric}) is shown
\cite{Koutsoumbas02}.
For large $\kappa$, A second-order phase transition occurs.
The criticality has been arousing much attention.
(b) 
For an extended parameter space, Eq. (\ref{extended_gonihedric}),
there emerge rich phases accompanying a multi-critical point
\cite{Hellmann93};
here, the self-avoidance parameter $\kappa$
is fixed 
($\kappa=1$).
In terms of this extended parameter space,
the transition point in Fig. \ref{figure1} (a) is identified
with the multi-critical point at $j=-0.25$.
}
\end{figure}

Because of the absence of the surface tension,
thermally activated undulations should be promoted
significantly.
Such feature might be reflected by the phase diagram;
see Fig. \ref{figure1} (a) \cite{Koutsoumbas02,Espriu97}.
We notice that 
a phase transition occurs at a considerably low temperature
quite reminiscent of that of
the two-dimensional Ising model.
Moreover, for large $\kappa$, the phase transition becomes a continuous one, whose criticality 
has been arousing much attention:
By means of the Monte Carlo method,
Johnston and Malmini \cite{Johnston96} obtained
the critical exponents $\nu=1.2(1)$, $\gamma=1.60(2)$ and $\beta=0.12(1)$
for the self-avoidance $\kappa=1$.
(Here, we quoted one typical set of exponents among those 
reported in the literature by various means.)
The authors claimed that the exponents bear remembrance to those
of the two-dimensional Ising model, namely, 
$\nu=1$, $\gamma=7/4$ and $\beta=1/8$.
On the other hand, with the Monte Carlo method,
Baig and co-worker \cite{Baig97}
obtained $\nu=0.44(2)$ ($\kappa=1$) and $\gamma/\nu=2.1(1)$ ($\kappa=0.5,1$).
By means of 
the low-temperature series expansion,
Pietig and Wegner \cite{Pietig98} obtained 
$\alpha=0.62(3)$, $\beta=0.040(2)$ and $\gamma=1.7(2)$ ($\kappa=1$).
With use of the cluster-variation method with the aid of the Pad\'e approximation
\cite{Cirillo97a,Cirillo97b,Cirillo98,Cirillo99},
Cirillo and co-worker obtained the
estimates $\beta=0.062(3)$ and
$\gamma=1.41(2)$.

Meanwhile, a subtlety of the Monte Carlo simulation
inherent to the gonihedric model
was noted by
Hellmann and co-worker \cite{Hellmann93}.
According to them, the relaxation to the thermal 
equilibrium is extremely slow,
and such slow relaxation smears out the singularity of the phase transition.
In order to cope with such slow relaxation (long auto-correlation length),
they employed the histogram Monte Carlo method.
However, the singularity of the phase transition could not be resolved satisfactorily.
(See also Ref. \cite{Cirillo98} for an alternative evidence of strong metastabilities.)
As a matter of fact, 
the gonihedric model at $\kappa=0$ reduces to the so-called
ferromagnetic $p$-spin model, 
and the model has been studied
extensively as a possible lattice realization of super-cooled liquids
and glassy behaviors \cite{Shore91,Swift00,Lipowski97,Lipowski00a,Lipowski00b,Lipowski00c}.
In this sense, an alternative simulation scheme other than the Monte Carlo
method is desirable in order to surmount the slow-relaxation problem
and determine the critical exponents reliably.

In this paper, we develop a transfer-matrix formalism, implementing
Novotny's idea \cite{Novotny90,Novotny92,Novotny93,Novotny91}, which enables us to treat arbitrary
number of spins $N$ for one transfer-matrix slice.
This arbitrariness admits systematic 
finite-size-scaling analyses.
(In addition to this advantage, 
the transfer-matrix calculation yields 
the correlation length $\xi$ directly.
Because $\xi$ has a fixed scaling dimension,
succeeding effort at the finite-size-scaling analyses
is reduced to a considerable extent.)
We also accept the idea of Cirillo and co-worker \cite{Cirillo97b},
who extended the parameter space of the gonihedric model (\ref{gonihedric}) to,
\begin{equation}
\label{extended_gonihedric}
J_1=-1, \   J_2=-j,  \  and\  J_3=-\frac{1-\kappa}{4\kappa}
   .
\end{equation}
(Note that for $j=-0.25$, the parameter space reduces to that of the aforementioned
original gonihedric model.)
With respect to this extended parameter space, Cirillo and co-worker
claimed 
that the above mentioned peculiar criticality
could be identified with a mere
end-point singularity (multi-criticality) \cite{Riedel69,Pfeuty74}
of an ordinary critical line of the three-dimensional-Ising
universality class;
see the critical branch of the phase diagram shown in 
Fig. \ref{figure1} (b).
Thereby,
they obtained the crossover critical
exponent $\phi=1.1(1)$ by means of the cluster-variation method \cite{Cirillo97b}.
Our transfer-matrix simulation supports their idea that the numerical data are
well described by the multi-critical (crossover) scaling theory.
We estimate the crossover exponent and the correlation-length critical exponent
as $\phi=0.6(2)$ and $\dot\nu=0.45(15)$, respectively;
hereafter, we place a dot over the critical indices at the multi-critical point ($j=-0.25$).

The rest of this paper is organized as follows.
In Sec. \ref{section2},
we set up a transfer-matrix formalism for the gonihedric model based on Novotny's idea.
In Sec. \ref{section3}, we present the numerical results.
Taking the advantage of the Novotny formalism, we carry out systematic finite-size-scaling analyses.
In the last section, we present summary and discussions.

\section{\label{section2}
Extension of the Novotny method to the plaquette-type interactions}

In this section, we present methodological details of 
our numerical simulation for the gonihedric model (\ref{gonihedric}).
We employed Novotny's improved version \cite{Novotny90,Novotny92,Novotny93} 
of the transfer-matrix method.
This technique allows us to construct the transfer matrices
containing arbitrary number of spins $N$ in one transfer-matrix slice;
note that in the conventional scheme, the available system sizes $N$
are limited for high spatial dimensions $d \ge 3$ severely.
Actually, Novotny constructed the transfer matrices of the Ising model
for $d \le 7$ fairly systematically \cite{Novotny91}.
Such arbitrariness of $N$ admits
systematic finite-size-scaling analyses.

In the following, we adopt Novotny's idea to study the gonihedric model
(\ref{gonihedric}).
For that purpose, we extend his idea so as to incorporate
plaquette-based interactions.
We restrict ourselves to the case of three dimensions $d=3$
relevant to our concern.
(The original idea of Novotny is formulated systematically for
general dimensions, taking the advantage that only the
bond-based (nearest neighbor) interaction is involved.)

\begin{figure}
\includegraphics{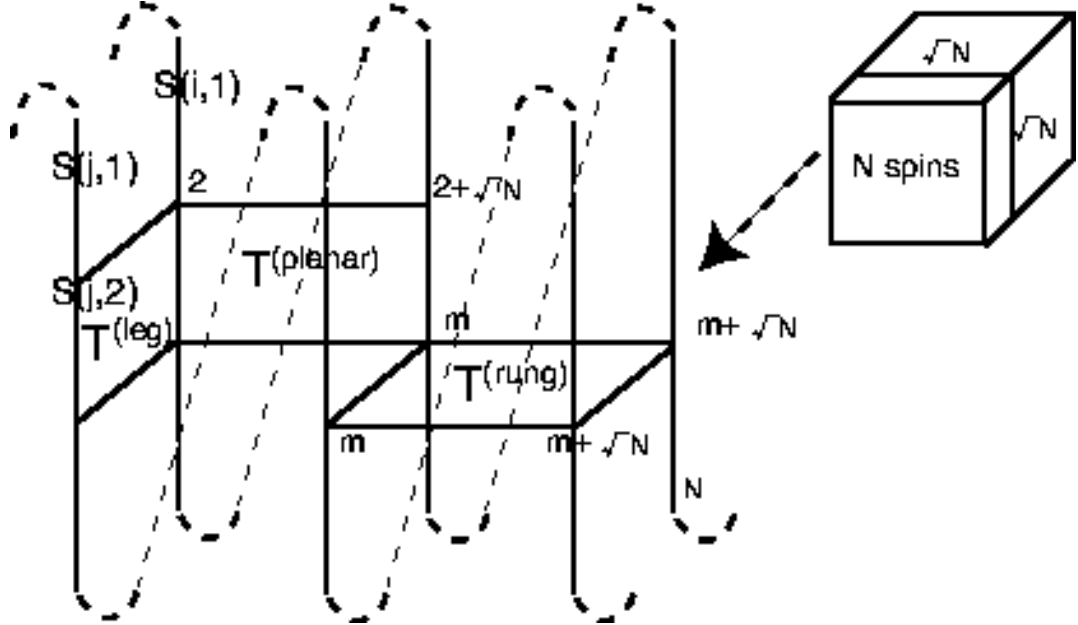}%
\caption{\label{figure2}
Novotny invented a new scheme to construct the
transfer matrix (TM) \cite{Novotny90,Novotny92,Novotny93},
which allows us to treat 
arbitrary number of spins $N$ per one TM slice.
We extend his scheme to incorporate the plaquette-type 
(next-nearest-neighbor and four-spin)
interactions, aiming to treat the gonihedric model (\ref{gonihedric}).
The contributions from the ``leg,'' ``planar,'' and ``rung''
interactions are considered separately; see Eq. (\ref{decomposed_TM}).
With use of the translation operator $P^{\sqrt{N}}$,
we build a bridge between the $\sqrt{N}$-th neighbor spins
along the leg (inter-leg interaction).
}
\end{figure}

We decompose the
transfer matrix into the following three components,
\begin{equation}
\label{decomposed_TM}
T=  T^{(leg)} \odot T^{(planar)} \odot T^{(rung)}     ,
\end{equation}
where the symbol $\odot$ denotes the Hadamard (element by element) matrix multiplication.
Note that the multiplication of the local Boltzmann weights
should give rise to the total Boltzmann factor.
The decomposed parts, $T^{(leg)}$, $T^{(planar)}$, and $T^{(rung)}$,
of Eq. (\ref{decomposed_TM}) stand for the Boltzmann weights for
intra-leg plaquettes, intra-planar plaquettes,
and rung plaquettes, respectively; see Fig. \ref{figure2} as well.

First, let us consider the contribution of $T^{(leg)}$.
The matrix elements are given by the formula,
\begin{equation}
T_{ij}^{(leg)}  =  \langle i| A | j \rangle
 = W_{S(i,1)S(i,2)}^{S(j,1)S(j,2)}
W_{S(i,2)S(i,3)}^{S(j,2)S(j,3)}
  \dots
W_{S(i,N)S(i,1)}^{S(j,N)S(j,1)}
   ,
\end{equation}
where the indices $i$ and $j$ specify the spin configurations
for both sides of the transfer-matrix slice.
More specifically, 
we consider $N$ spins for a transfer-matrix slice,
and
the index $i$ specifies a spin configuration 
$\{ S(i,1),S(i,2), \dots , S(i,N) \}$
arranged along the leg;
see Fig. \ref{figure2}.
The factor $W_{S_1 S_2}^{S_3 S_4}$ denotes the local Boltzmann weight for 
a plaquette with corner spins $\{ S_1, \dots ,S_4 \}$.
Explicitly, it is given by the following form,
\begin{equation}
W_{S_1 S_2}^{S_3 S_4} =
\exp \left( - \frac{1}{T}\left(
   \frac{J_1}{4}(S_1 S_2 +S_2 S_4+S_4 S_3+S_3 S_1)
 + \frac{J_2}{2}(S_1 S_4 + S_2 S_3)
 + \frac{J_3}{2}S_1 S_2 S_3 S_4   \right)  \right)    
   .
\end{equation}
(The denominators of the coupling constants are intended to avoid double counting.)
Here, the parameter $T$ denotes the temperature.
It is to be noted that the component $T^{(leg)}$,
with the other components ignored, 
leads the transfer-matrix for the two-dimensional gonihedric model.
The other components of $T^{(planar)}$ and $T^{(rung)}$ should
introduce the ``inter leg'' interactions so as to raise the
dimensionality to $d=3$.

Second,
we consider the component for the intra-planar interaction.
It is constructed by the following formula,
\begin{equation}
T_{ij}^{(planar)} = \langle i| A P^{\sqrt{N}} |i \rangle  ,
\end{equation}
where the matrix $P$ denotes the translation operator;
namely, with the operation,
a spin arrangement $\{ S(i,m) \}$ is shifted to $\{ S(i,m+1) \}$;
the periodic boundary condition is imposed.
An
explicit representation of $P$ is given afterward.
Because of the insertion of $P^{\sqrt{N}}$, the plaquette interaction $A$ 
bridges the $\sqrt{N}$-th-nearest-neighbor pairs,
and so, it brings about the desired inter-leg interactions.
This is an essential idea of Novotny's work.
Crucial point is that 
the operation $P^{\sqrt{N}}$
is still meaningful, even though the power $\sqrt{N}$ is an irrational number
\cite{Novotny90,Novotny92,Novotny93}.
This rather remarkable fact renders freedom that
one can choose arbitrary number of spins.

An explicit representation of $P^x$ is given as follows \cite{Novotny90,Novotny92,Novotny93}.
As is well known, the eigenvalues $\{ p_k \}$ 
of $P$ belong to the $N$ roots of unity 
like
$\exp(i\phi_k)$ with $\phi_k=2\pi k/N$ ($k=0,1,\dots,N-1$).
The complete set of the corresponding eigenvectors are
constructed by the formula
$ |\Phi_k\rangle = 
    N_{\Phi_k}^{-1} \sum_{l=1}^{N} p_k^l P^l | \Phi  \rangle$.
Here, the set  $\{ | \Phi \rangle \}$ consists of
such bases independent
with respect to the translation operations, 
and $N_{\Phi k}^{-1}$ is a normalization factor.
Provided that the eigenstates $| \Phi_k \rangle$ are at hand,
one arrives at an explicit representation of $P^x$;
\begin{equation}
\langle i | P^x | j \rangle = \sum_{\Phi_k}  \langle i | \Phi_k\rangle 
                                p_k^x \langle \Phi_k | j \rangle .
\end{equation}

Finally, we consider
the component of $T^{(rung)}$.
This component is also constructed similarly.
This time, however, we need two operations of $P^{\sqrt{N}}$,
because $T^{(rung)}$ concerns both sectors of $i$ and $j$
(both sides of the transfer-matrix slice);
see Fig. \ref{figure2}.
The elements are given by,
\begin{equation}
T^{(rung)}_{ij} =
  \left( \langle i| \otimes \langle j| \right) B 
      \left( \left(P^{\sqrt{N}} |i\rangle\right) 
   \otimes \left(P^{\sqrt{N}} |j\rangle\right) \right)  ,
\end{equation}
where the operator $B$ acts on the direct-product space;
\begin{equation}
 \left(  \langle i| \otimes \langle j| \right) B 
    \left(  |k\rangle \otimes |l\rangle \right) = 
  \prod_{m=1}^{N} W_{S(i,m) S(j,m)}^{S(k,m) S(l,m)} .
\end{equation}
Putting the components, $T^{(leg)}$, $T^{(planar)}$ and $T^{(rung)}$,
 into Eq. (\ref{decomposed_TM}), 
we obtain the complete form of the transfer matrix.
Actual numerical diagonalizations are performed
in the following section.

\section{\label{section3}
Numerical results}

In this section, we survey the criticality of the gonihedric model
(\ref{gonihedric}) for the extended parameter space (\ref{extended_gonihedric})
by means of the transfer-matrix method developed in the preceding section.
In particular,
we investigate the critical branch with an emphasis on the end-point singularity at $j=-0.25$.
We neglect a possible deviation of the multi-critical point
from $j=-0.25$ as pointed out by the cluster-variation-method study 
\cite{Cirillo97b}.
Such deviation is so slight that it would not affect
the multi-critical analyses
very seriously \cite{Cirillo97b}.
We treated the system sizes
up to $N=13$.
The system sizes $N$ are restricted to odd numbers,
for which the transfer-matrix elements consist of real numbers 
\cite{Novotny90,Novotny92,Novotny93}.

\subsection{
Survey of the critical branch with the Roomany-Wyld approximative beta function}

To begin with, we survey the criticality of the second-order phase boundary
in Fig. \ref{figure1} (b).
For that purpose, we calculated the Roomany-Wyld approximative
beta function $\beta^{RW}(T)$.
We stress that the availability of $\beta^{RW}(T)$ is one of
major advantages of the transfer-matrix method.
The Roomany-Wyld beta function is given by the following formula \cite{Roomany80}, 
\begin{equation}
\label{RW_beta_function}
\beta^{RW}_{N}(T) =
  - \frac{1-  \frac{\ln(\xi_N(T)/\xi_{N-2}(T))}
              {\ln(\sqrt{N}/\sqrt{N-2})} }
        {\sqrt{   \frac{\partial_T \xi_N(T) \partial_T \xi_{N-2}(T)}
                        {\xi_N(T) \xi_{N-2}(T) }  }   }   .
\end{equation}
Here, $\xi_N(T)$ denotes the correlation length for the system size $N$,
The correlation length is readily calculated by means of the transfer-matrix
method.
That is,
using the largest and next-largest eigenvalues, namely,
$\lambda_1$ and $\lambda_2$, of the transfer matrix,
we obtain the correlation length
$\xi=1/ \ln (\lambda_1 / \lambda_2)$ immediately.

In Fig. \ref{figure3}, we plotted the beta function $\beta^{RW}_{13}(T)$ for various
$j$ with the fixed self-avoidance parameter $\kappa=2$.
The zero point (fixed point) of the beta function $\beta^{RW}_{13}(T)$
indicates the location of the critical point
$T_c$.
In Inset of Fig. \ref{figure3}, we plotted the phase-transition point $T_c(j)$.
This phase boundary corresponds to the critical branch of the phase
diagram shown in Fig. \ref{figure1} (b); the other phase boundaries are of first order,
and the determination of them is out of the scope of 
the present $\beta^{RW}_N(T)$ approach.

The slope of the beta function at $T=T_c$ yields an estimate for the inverse of the
correlation-length critical exponent $1/\nu$.
In Fig. \ref{figure3},
we also presented a slope (dashed line) 
corresponding to the three-dimensional-Ising universality
class $\nu=0.6294$ \cite{Ferrenberg91} for a comparison.
We see that the criticality is maintained to be the three-dimensional-Ising universality class
for a wide range of $j$.
More specifically, for $j=-0.05$, $0.1$, $0.25$, $0.4$, $0.55$, and
$0.7$, we obtained the correlation-length critical exponent as
$\nu=0.634$, $0.641$,
$0.643$, $0.643$, $0.642$, and $0.642$, respectively.
From this observation, we estimated the exponent along the
critical branch as $\nu=0.638(5)$ fairly in good agreement
with
the three-dimensional-Ising universality class.

It is to be noted that, as mentioned in Introduction,
at $j=-0.25$, very peculiar critical exponents have been reported so far 
\cite{Johnston96,Baig97,Pietig98,Cirillo97b}.
The above simulation result suggests that such 
peculiar criticality should be realized only at
$j=-0.25$ (critical end-point).
This idea was first claimed by Ref. \cite{Cirillo97b} with the
cluster-variation method.
In fact, on closer inspection, the beta function in Fig. \ref{figure3} shows a crossover behavior
such that the slope in the off-critical regime is enhanced;
see the regime 
of $T-T_c > 3$ at $j=-0.05$ in particular.
It appears that
such regime of enhancement is pronounced as $j \to -0.25$.
Eventually,
right at $j=-0.25$, a new universality accompanying small $\dot\nu(<\nu)$ may emerge.
In the succeeding subsections, we provide further support to this issue.

For the region in close vicinity to the critical end-point,
for instance, $-0.25 < j<-0.2$,
we found that the beta function acquires unsystematic finite-size corrections;
even the zero point of $\beta^{RW}_N(T)$ disappears.
In this sense, we suspect that a direct simulation at $j=-0.25$ would not be
very efficient.
Rather, performing simulations for a wide range of $j$,
we are able to extract informations concerning the end-point singularity
fairly reliably.

\subsection{
End-point
singularity of the critical amplitude of $\xi$}

In the above, we found that the universality class of the critical branch
is maintained to be that of the three-dimensional Ising model.
A notable feature is that a crossover to a new universality class
emerges as $j \to -0.25$ .
In this subsection, we study this multi-criticality 
in terms of the theory of the crossover critical phenomenon.
We read off the crossover exponent $\phi$
from the end-point singularity of the amplitude \cite{Pfeuty74}
of the correlation length.
Namely, the correlation length should diverge in the form,
\begin{equation}
\xi \approx N^{\pm} |T-T_c|^{-\nu}, 
\end{equation}
with the amplitude,
\begin{equation}
\label{singurality_xi}
   N^\pm \propto \Delta^{(-\dot{\nu}+\nu)/\phi}   .
\end{equation}
Here, the variable $\Delta$ denotes the distance from the
multi-critical point $\Delta=j+0.25$.
(It is to be noted that the critical point
$T_c$ depends on $\Delta$ as demonstrated in 
Inset of Fig. \ref{figure3}.)
The above formula is a straightforward consequence of the multi-critical (crossover) scaling
hypothesis \cite{Riedel69,Pfeuty74};
\begin{equation}
\label{cross-over_scaling}
\xi \approx |T-T_c|^{-\dot{\nu}} X(\Delta/|T-T_c|^\phi)  .
\end{equation}
As noted in the previous subsection,
the dotted critical index stands for that right at the multi-critical point.

To begin with, we determine the critical amplitude $N^+$.
In Fig. \ref{figure4}, we plotted the scaled correlation length
$(T-T_c)L^{1/\nu}$-$\xi |T-T_c|^\nu$ 
for $\kappa=2$ and $j=0.3$.
The symbols, $+$, $\times$, $*$, $\Box$, and $\blacksquare$,
denote the system sizes of $N=5$, $7$, $9$, $11$, and $13$,
respectively.
The linear dimension of the system $L$ is given by $L=\sqrt{N}$.
In the plot, we postulated the three-dimensional-Ising universality class $\nu=0.6294$ 
\cite{Ferrenberg91}.
We see that the scaled data collapse into a scaling-function curve.
We again confirm that the phase transition belongs to the
three-dimensional-Ising universality class.
In addition to this,
from the limiting value of the high-temperature side of
the scaling function, 
we estimate the critical
amplitude as $N^+=2.09(13)$ for $\kappa=2$ and $j=0.3$;
more specifically,
we read off the value of $N=13$ around the regime $(T-T_c)L^{1/\nu} \approx 30$,
and
as for an error indicator,
we accepted the amount of the data scatter among $N=5,\dots,13$.

Similarly, we determined $N^+$ for various parameter ranges of both
$j$ and $\kappa$.
In Fig. \ref{figure5}, we plotted the amplitude $N^+$ for 
$\kappa=1$, $2$ and $4$ with 
$\Delta(=j+0.25)$ varied.
In the plot, we observe a clear signature of the 
power-law singularity as described by Eq. (\ref{singurality_xi}).
Hence, we confirm 
that the cross-over behavior (\ref{cross-over_scaling})
is realized actually around the multi-critical point $j=-0.25$.
Moreover, in the figure, we notice that the data for $\kappa=1$, $2$ and $4$
almost overlap each other.
It would be rather remarkable that the amplitude $N^+$ itself
hardly depend on the parameter
$\kappa$.
This fact indicates that the multi-criticality, namely, the singularity exponent
$(-\dot\nu+\nu)/\phi$, stays universal
with respect to the self-avoidance parameter $\kappa$.
Such universality was first reported by the series-expansion analyses
surveying the range of
$\kappa=0.5,\dots,3$ \cite{Pietig98}.

From the slopes in Fig. \ref{figure5},
we obtained the
singularity exponent as $(-\dot{\nu}+\nu)/\phi=0.422(6)$, $0.405(5)$. and $0.415(7)$
for $\kappa=1$, $2$ and $4$, respectively.
We estimate the singularity exponent as
$(-\dot{\nu}+\nu)/\phi=0.415(20)$ consequently.

Let us mention some remarks on this estimate
$(-\dot{\nu}+\nu)/\phi=0.415(20)$.
First, this result excludes such a possibility $\dot{\nu} > \nu $
as $\dot{\nu}=1.2(1)$ \cite{Johnston96}.
Rather, our result supports the results of
$\nu=0.44(2)$ ($\kappa=1$) with the Monte Carlo method \cite{Baig97}
and $\nu=0.46(1)$ ($\kappa=1$)
with the low-temperature-series-expansion result \cite{Pietig98}.
(The latter is obtained from $\dot\alpha=0.62(3)$ ($\kappa=1$) \cite{Pietig98}
together with the hyperscaling relation $\dot\alpha = 2- d \dot\nu$.)
Note that
our preliminary survey in the preceding subsection
also indicates a signature of $\dot{\nu} < \nu $.

Second, postulating the value $\dot\nu \approx 0.45$ 
close to the aforementioned existing values,
we obtain an estimate for the crossover exponent $\phi \approx 0.43$.
The present result contradicts the result $\phi=1.1(1)$ \cite{Cirillo97b}
determined with the cluster-variation method.
In the succeeding section, we will provide further support to
$\phi \approx 0.43$, performing the multi-critical scaling analysis
based on the relation (\ref{cross-over_scaling}).

\subsection{Multi-critical scaling analysis}

In the above, we obtained an estimate for the crossover
exponent $\phi \approx 0.43$ from the power-law singularity of the amplitude $N^+$,
accepting the value $\dot\nu \approx 0.45$ advocated by Refs. \cite{Baig97,Pietig98}.
In this subsection, we provide further support to these exponents.
We
carry out an multi-critical (crossover) scaling analysis based on Eq. (\ref{cross-over_scaling}).
For finite size $L$, the scaling-hypothesis formula should be extended to,
\begin{equation} 
\label{cross-over_scaling_L}
\xi=L\tilde{X}((T-T_c)L^{1/\dot\nu},\Delta L^{\phi/\dot\nu}).
\end{equation}
Based on this formula,
in Fig. \ref{figure6}, we present the scaled data,
$(T-T_c) L^{1/\dot{\nu}}$-$\xi/L$,
with fixed $\Delta L^{\phi/\dot{\nu}} = 2$ and $\kappa=2$.
Here, we set the exponents $\dot\nu=0.4$ and $\phi=0.6$ for which we found the best data collapse.
Surveying the parameter space beside this condition, we obtained the critical exponents
as $\dot\nu=0.45(15)$ and $\phi=0.6(2)$.
These estimates agree with the analysis in the preceding subsection.

We stress that the use of $\xi$ greatly simplifies the scaling analyses,
because $\xi$ has a {\em fixed} scaling dimension, namely, [length]${}^{1}$.
For instance, as for other quantities such as  
the susceptibility, we need to determine the exponent
$\dot{\gamma}$ in addition to $\dot{\nu}$.
In this sense, the present approach via the transfer matrix
is advantageous over other approaches.

\section{Summary and discussions}

We investigated the (multi) criticality of the gonihedric model 
(\ref{gonihedric}) with the extended parameter space
(\ref{extended_gonihedric}).
The model is notorious for its slow relaxation to the thermal equilibrium
(glassy behavior), 
which deteriorates the efficiency of the Monte Carlo sampling \cite{Hellmann93}.
Aiming to surmount the difficulty,
we employed the transfer-matrix method.
We implemented Novotny's idea \cite{Novotny90,Novotny92,Novotny93},
extending it so as to incorporate the plaquette-type interactions 
(Sec. \ref{section2}).
The present approach enables us to treat arbitrary number of spins
per one transfer-matrix slice, admitting systematic 
finite-size-scaling analyses; see Fig. \ref{figure4} for instance.

\begin{figure}
\includegraphics{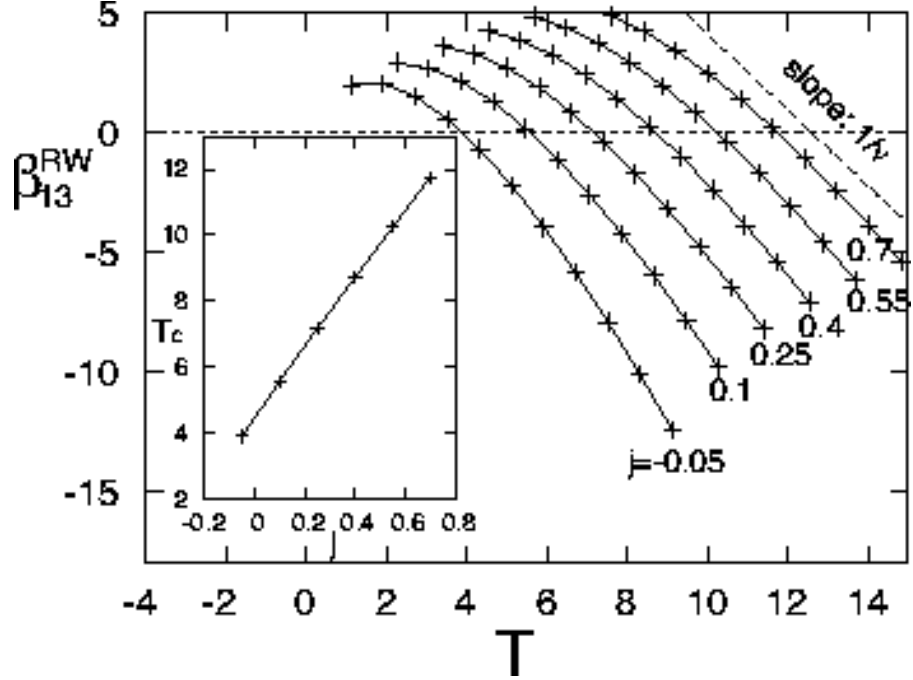}%
\caption{\label{figure3}
The beta function $\beta^{RW}_{13}(T)$ (\ref{RW_beta_function}) 
is plotted for $\kappa=2$ and various $j$.
For a comparison, we presented a slope (dashed line)
corresponding to the three-dimensional-Ising 
universality class
($\nu=0.6294$ \cite{Ferrenberg91});
we see that the criticality is maintained to be the three-dimensional-Ising 
universality class for a wide range of $j$.
In fact,
from the slopes at the fixed points of $\beta^{RW}_{13}(T)$, we obtain an estimate
for the correlation-length critical exponent
$\nu=0.638(5)$; see text for details.
Inset: Plotting the zero-points of $\beta^{RW}_{13}(T)$, we determine
a phase boundary $T_c(j)$, which corresponds to the critical branch in 
Fig. \ref{figure1} (b).
}
\end{figure}

\begin{figure}
\includegraphics{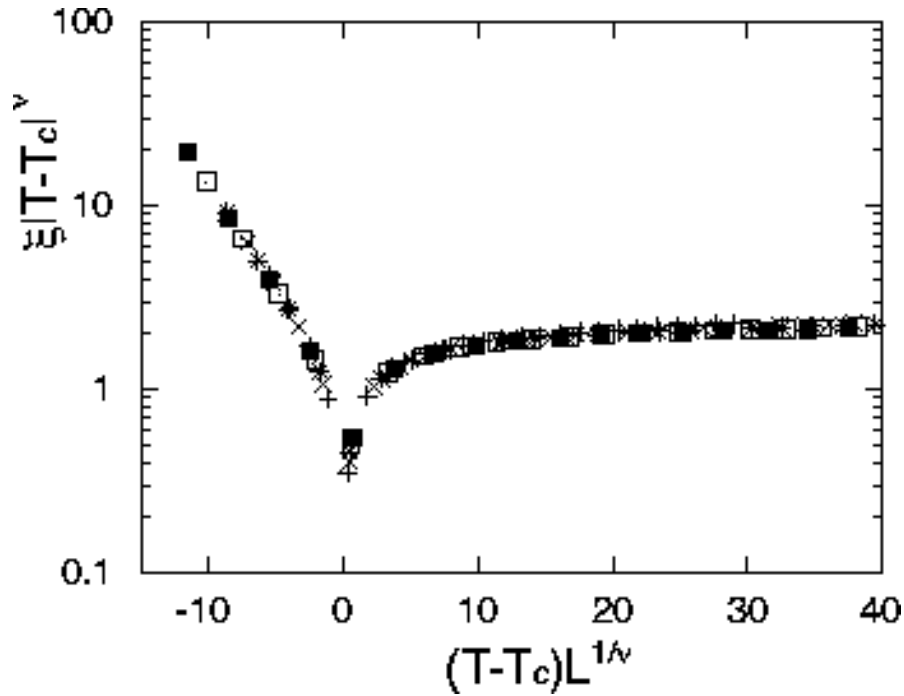}%
\caption{\label{figure4}
Scaling plot for the correlation length, namely,
$(T-T_c)L^{1/\nu}$-$\xi |T-T_c|^\nu$, is shown
for
$\kappa=2$ and $j=0.3$.
Here, we postulated the three-dimensional-Ising
universality class $\nu=0.6294$ \cite{Ferrenberg91}.
The symbols, $+$, $\times$, $*$, $\Box$, and $\blacksquare$,
denote the system sizes of $N=5$, $7$, $9$, $11$, and $13$,
respectively.
We confirm that the transition belongs to the three-dimensional-Ising universality class.
Furthermore,
from the plateau in the high-temperature side, we obtain
an estimate for the critical
amplitude $N^+=2.09(13)$; see text for details.
}
\end{figure}

The transfer-matrix calculation has an advantage
in that it yields the correlation length immediately.
Because the correlation length has a known (fixed) scaling dimension,
the subsequent scaling analyses are simplified significantly.
Moreover, with the correlation length,
we are able to calculate 
the Roomany-Wyld approximate beta function 
$\beta^{RW}_N(T)$ (\ref{RW_beta_function}).
With use of $\beta^{RW}_{N}(T)$,
we surveyed the critical branch of the phase diagram (Fig. \ref{figure3}).
Thereby, we observed that the criticality is maintained to be
the three-dimensional-Ising universality class 
all along the phase boundary.
On closer inspection, we found an indication of a crossover critical phenomenon
such that the slope of $\beta^{RW}_{13}(T)$ in the off-critical regime,
typically, $T-T_c>3$ ($j=-0.05$), acquires a notable enhancement.
This fact indicates that a multi-criticality with smaller $\dot\nu$
emerges as $j\to-0.25$.
This observation supports the claim \cite{Cirillo97b}
that the nonstandard criticality reported so far
\cite{Johnston96,Baig97,Pietig98,Cirillo97a,Cirillo97b,Cirillo98}
could be attributed to the
end-point criticality specific to $j=-0.25$.

\begin{figure}
\includegraphics{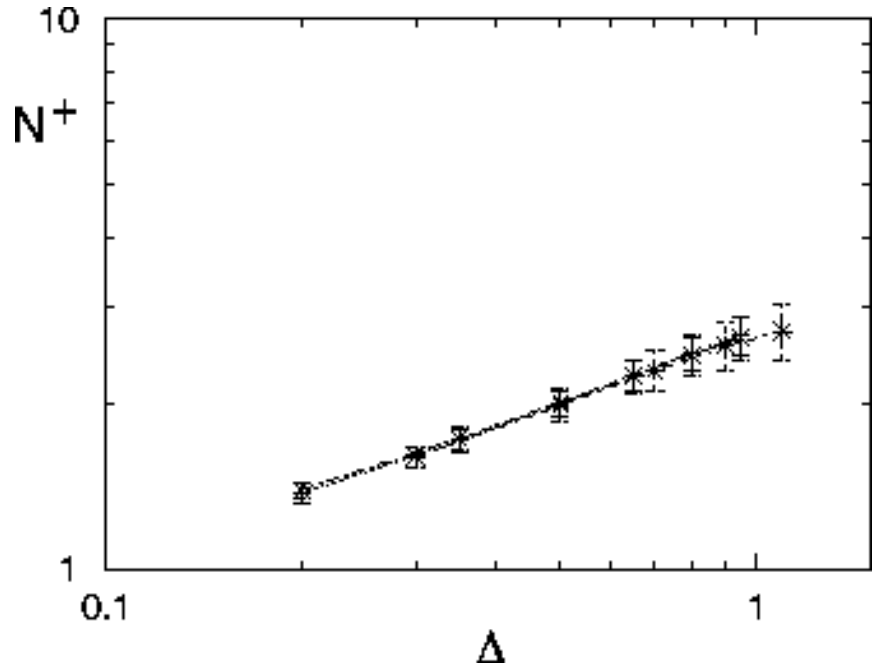}%
\caption{\label{figure5}
Correlation-length 
critical amplitude $N^+$ (\ref{singurality_xi})
is plotted for various $\Delta(=j+0.25)$ and 
$\kappa=0.5$, $1$ and $4$.
The symbols, $+$, $\times$, and $*$,
stand for the self-avoidance parameter $\kappa=0.5$, $1$, and $4$,
respectively.
The data indicate a clear power-law singularity, Eq. (\ref{singurality_xi}).
From the slopes, we estimate the singularity exponent as
$(-\dot{\nu}+\nu)/\phi=0.415(20)$.
}
\end{figure}

Aiming to clarify the nature of this multi-criticality,
we analyzed the end-point singularity of the amplitude 
of the correlation length $N^+$ (\ref{singurality_xi}).
As shown in Fig. \ref{figure5},
the amplitude exhibits a clear power-law singularity, 
from which we
obtained an estimate for the singularity exponent $(-\dot{\nu} + \nu)/\phi=0.415(20)$.
This result supports the above-mentioned observation that an inequality
$\dot\nu < \nu$ should hold,
and in other words,
it excludes such a possibility of $\dot{\nu} > \nu$ 
advocated in Ref. \cite{Johnston96}.
Rather, our result supports
the Monte Carlo simulation result 
$\dot\nu=0.44(2)$ ($\kappa=1$) \cite{Baig97}
and the low-temperature-series-expansion result $\dot\nu=0.46(1)$ ($\kappa=1$)
\cite{Pietig98}.
Postulating $\dot\nu \approx0.45$, 
we arrive at an estimate for the crossover exponent $\phi \approx 0.43$.
This exponent is to be compared with the result 
$\phi=1.1(1)$ determined with the cluster-variation method \cite{Cirillo97b}.
The discrepancy between the result \cite{Cirillo97b} and ours seems to be rather conspicuous.

\begin{figure}
\includegraphics{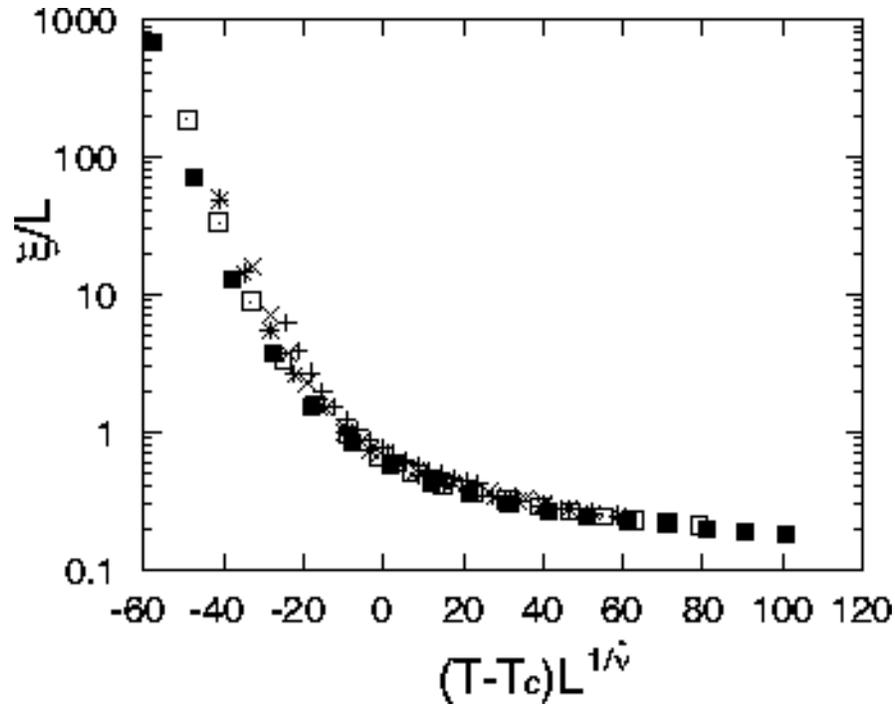}%
\caption{\label{figure6}
Multi-critical (crossover) scaling plot (\ref{cross-over_scaling_L}),
$(T-T_c) L^{1/\dot{\nu}}$-$\xi / L$, for 
$\kappa=2$ and $\Delta L^{\phi/\dot\nu}=2$ is shown.
Here, we set $\dot\nu=0.4$ and $\phi=0.6$,
 for which we found the 
best data collapse.
The symbols, $+$, $\times$, $*$, $\Box$, and $\blacksquare$,
denote the system sizes of $N=5$, $7$, $9$, $11$, and $13$,
respectively.
}
\end{figure}

We then carried out the multi-critical scaling analysis (\ref{cross-over_scaling_L})
in order to provide further support to our estimate 
$\phi \approx 0.43$ based on $\dot\nu \approx 0.45$.
We found that a good data collapse is attained
for $\dot\nu=0.4$ and $\phi=0.6$ under $\kappa=2$ and $\Delta L^{\phi/\dot\nu}=2$
(Fig. \ref{figure6}).
Surveying the parameter space,
we obtained the estimates $\phi=0.6(2)$ and $\dot\nu=0.45(15)$.
These exponents agree with the above-mentioned analysis via
the critical amplitude $N^+$.

As a consequence, we confirm that the whole analyses 
managed in this paper
lead a self-consistent conclusion.
Regarding the discrepancy on $\phi$,
we suspect that the value $\phi=1.1(1)$ \cite{Cirillo97b}
might be rather inconceivable.
Nevertheless,
in order to fix the multi-criticality more definitely,
further elaborate investigations would be required.
As a matter of fact,
a possible slight deviation of the multi-critical point from $j=-0.25$
was ignored throughout the present work as in Ref. \cite{Cirillo97b}.
Justification of such a treatment might be desirable.
In any case,
the present approach, which is completely free from
the slow-relaxation problem, would provide a promising
candidate for a 
first-principles-simulation scheme in future research.

\begin{acknowledgments}
This work is supported by Grant-in-Aid for
Young Scientists
(No. 15740238) from Monbu-kagakusho, Japan.
\end{acknowledgments}


\end{document}